\documentclass[aps,prd,superscriptaddress,onecolumn,groupedaddress,amssymb,nofootinbib,10pt]{revtex4}
\pdfoutput=1
\usepackage[pdftex]{graphicx}
\DeclareGraphicsExtensions{.pdf,.jpeg,.png}
\usepackage{bm}
\usepackage{color}
\usepackage{dcolumn}
\usepackage[spanish,english]{babel}
\usepackage{latexsym,float}
\usepackage{epsfig,graphics}
\usepackage{epstopdf}
\usepackage{amssymb}
\usepackage{mathtools}
\usepackage{amsmath}
\usepackage{verbatim}
\usepackage{multirow}
\usepackage{color}
\usepackage{tikz}
\usepackage[utf8]{inputenc}
\usetikzlibrary{matrix}
\usepackage{float}

\usepackage{ulem}
\def\clap#1{\hbox to 0pt{\hss#1\hss}}

\def\bea{\begin{eqnarray}}
\def\eea{\end{eqnarray}}
\def\be{\begin{equation}}
\def\ee{\end{equation}}

\renewcommand{\geq}{\geqslant}
\renewcommand{\leq}{\leqslant}

\begin{document}

\title{Chaos in balanced and unbalanced holographic s+p superconductorss }
\author{Davood Momeni} \email{davood@squ.edu.om}
\affiliation{Center for Space Research, North-West University, Mafikeng, South Africa }
\affiliation{Tomsk State Pedagogical University, TSPU, 634061 Tomsk, Russia } \affiliation{Department of Physics, College of Science, Sultan Qaboos University,\\P.O. Box 36, ,AL-Khodh 123  Muscat, Sultanate of Oman \footnote{Permanent address}
}

\author{Nayereh Majd} \email{naymajd@ut.ac.ir}
\affiliation{{Department of Engineering Science, Faculty of Engineering, University of Tehran, Tehran, PO Box 11155-4563, Iran}}

\author{Morteza Mohammadzaheri}\email{mmzahery@gmail.com}
\affiliation{Department of Mechanical \& Industrial Engineering, College of Engineering, Sultan Qaboos University, P.O. Box 33, Al-khodh 123 Muscat, Sultanate of Oman}

\author{Phongpichit Channuie}
\email{channuie@gmail.com}
\affiliation{ College of Graduate Studies, Walailak University, Thasala, \\Nakhon Si Thammarat, 80160, Thailand}
\affiliation{ School of Science, Walailak University, Thasala, \\Nakhon Si Thammarat, 80160, Thailand}

\author{Mudhahir Al Ajmi} \email{mudhahir@squ.edu.om}
\affiliation{{Department of Physics, College of Science, Sultan Qaboos University,P.O. Box 36, ,AL-Khodh 123  Muscat, Sultanate of Oman}}

\date{\today}

\begin{abstract}
In this work, we propose a toy model for mixture of superconductors with competitive s and p modes using gauge/gravity duality. We demonstrate that the model undergoes phase transitions with the proper choice of different values for chemical potentials. We consider both balanced and unbalanced cases. 
We  propose that the condensate field in the bulk toy model for a mixing s+p phase of high temperature superconductors undergoes a chaotic phase space scenario. Using a suitable measure function for chaos, we demonstrate the existence of chaotic dynamics via condensate fields. As an alternative method, we also investigate the phase portrait in the extended phase space for the condensates. 

\end{abstract}


\maketitle
\section{Introduction}
In the sense of the AdS/CFT correspondence \cite{maldacena}, any weakly gravitational model in the bulk has a dual description based on quantum field theory. The correspondence is used as a tool to map the theories from the strongly
coupled sector to an equivalently classical gravity.
Holographic superconductivity is one of the successful implications of the correspondence first proposed in Refs.\cite{HSC1,HSC2,HSC3}. 
From the symmetry point of view, spontaneous symmetry breaking of the holographic superconductors may result in different circumstances. Regarding the scalar condensates, the $U(1)$ symmetry is broken ($s$-wave phase). In addition, the vector condensation has a different feature ($p$-wave phase). Recently it has been shown that there is a possibility to have two $s$-wave phases with two different scalar condensates \cite{Musso:2013ija}. The study was also extended to include the case
of the spontaneously symmetry breaking under the rotational symmetry by a vector as a possibility for the coexistence of two $p$-wave order parameters \cite{Amoretti:2013oia}. However, an intermediate phase between $s$- and $p$-phases known as the $s+p$- phase is much compelling. In this phase we have simultaneously two kinds of the relevant operators on the boundary. The CFT description has a direct interpretation for $s+p$ as a candidate of a mixed phase. More precisely, this model is based on the $U(2)$ gauge fields and the numerical studies of the
the $s+p$ wave phase in a typical holographic
superconductor are in \cite{Nie:2013sda}. 

A toy model of the multi band holographic superconductors was first proposed in \cite{Krikun:2012yj} and it was further extended by the authors of Ref.\cite{Amado:2013xya}. Here the model is based on the AdS/CFT described by the two-component super fluidity and it has been shown that the system being under perturbation becomes unstable. Also, the AdS/CFT correspondence implies always chaos with classical bulk duals \cite{deBoer:2017xdk}.
More interestingly, allocating a chaos index for the dynamics of generic quantum field theories is a challenging problem. 
New research in chaotic motion
of classical strings in AdS-like spacetimes can be found in \cite{29,43}.
The chaotic behavior  of the chiral condensate of the linear $\sigma$ model of low-energy QCD was studied in Ref.\cite{1605.08124},
which serves as a toy model of chaos of a quantum phenomenon.
In addition, an $N$ = 2 supersymmetric
QCD with the $SU(N_c)$ gauge group
at large $N_c$ and at strong coupling \cite{44} has been studied.  
It is worth noting that the chaotic behavior can be globally observed in Nature, e.g., chaos in black holes \cite{12-12,13-13,14-14}. Recently, the authors of \cite{Wang:2016wcj} investigated a test scalar particle coupling to Einstein tensor in the Schwarzschild-Melvin black hole spacetime through the short-wave approximation. Here the effects of coupling parameter on the chaotic behavior of the tested particles was examined.

The Melnikov method is also an attractive tool in performing analytical studies of chaotic systems~\cite{24-24, 25-25}, based on the so called Melnikov integral. The key idea of Melnikov
method is to measure the distance between the invariant
manifolds along the homoclinic orbit of the unperturbed vector
field with an assumption of the time periodic
and small perturbation where the distance can be calculated using the Melnikov function. The condition in which the Melnikov function has a simple zero point, provides important constraints on the black hole charge that must be implemented to observe the chaotic features. Sufficiently small perturbations can generate chaotic behavior due to temporal thermal fluctuations. In other words, a small spatially periodic perturbation in the equilibrium configuration with an absolute temperature above critical temperature can produce special chaos in the equilibrium configuration. 

Note that the existence of chaotic bahavior in gauge/gravity duality is also known phenomena \cite{Luo:2017bno,Huang:2017gih,Polchinski:2016hrw,Mezei:2016wfz,Hashimoto:2016wme,Ge:2014aza,Farahi:2014lta,Basu:2013vva,Zhang:2012uy,Floratos:2011ct}. In addition, it has been shown that, in the Sachdev-Ye-Kitaev (SYK) model \cite{SYK}, the hydrodynamic description universally describes large $N$ systems with an emergent conformal invariance, and consequently features maximally chaotic behavior \cite{prl}. An important characteristic of chaos is sensitive to initial conditions implying a limitation of numerical predictions of its trajectory.
\par
In this work, we study balanced and unbalanced holographic $s+p$ superconductors. We pave our setup for the bulk action of a mixed phase $s+p$ holographic superconductor in Sec.\ref{2}. Here we derive a set of equations of motion for scalar and gauge fields. In Sec.\ref{3}, we consider the chaotic behavior by firstly focusing on the balanced holographic case. In addition, the unbalanced scenario will be subsequently investigated in Sec.\ref{4}. Finally, we conclude our findings in the last section. 

\section{Competitive modes: s+p phase}
\label{2}
A suitable bulk action for a mixed phase $s+p$ holographic superconductor can be built using an AdS background metric $g_{\mu\nu}$ with negative scalar curvature $R$ appended by a $U(2)$ gauge field with a field strength $F_{c}^{\mu \nu }$ where $c=1,2,3$ and $\mu,\nu=0,1,2,3$. In addition, a scalar field doublet $\Psi\in \mathcal{C}$ plays a role of the condensate field, proposed in \cite{Amado:2013lia}:
\begin{eqnarray}
S&=&\int d^{4}x\sqrt{-g}\Big(\frac{R-2\Lambda }{2\kappa ^{2}}\Big)+\int d^{4}x\sqrt{-g}%
\Big(-\frac{1}{4}F_{c}^{\mu \nu }F_{\mu \nu }^{c}-m^{2}|\Psi
|^{2}-|D^{\mu }\Psi |^{2}\Big).  \label{action}
\end{eqnarray}%
where $\kappa^2=8\pi G$ is gravitational coupling constant, $R$ is the Ricci scalar for the metric of AdS spacetime $R=g^{\mu\nu}R_{\mu\nu}$, and $g\equiv \det(g_{\mu\nu})$ is the determinant of the metric. The covariant derivative operator is commonly given as $%
D_{\mu }=\partial _{\mu }-iqA_{\mu }$ and $q$ is an $U(2)$ charge.  Here $\Lambda$ is the negative cosmological constant for the AdS spacetime, i.e. $\Lambda<0$, and  $m^{2}$ is the mass parameter for the scalar-doublet  above the Breitenlohner-Freedman (BF) bound  \cite{BF1,BF2}. The bulk
AdS metric with planar
topology of the horizon in the Schwarzschild coordinates system $(t,r,\theta,\varphi)$ can be written as follows:
\begin{equation}
g_{\mu \nu }={\rm diag}\Big(-f(r),\,f(r)^{-1},\,r^{2}\Sigma _{2}\Big)\,,  \label{metric}
\end{equation}%
with $\Sigma_{2} = {\rm diag}(1,\,\sin ^{2}\theta )$. The proposed model was intended to describe the mixture phase $s+p$ far from the probe limit. However, the back reaction effects will be omitted in this paper. This is called a holographic toy model. Regarding AdS/CFT manner, its boundary dual corresponds to a lower three-dimensional high temperature superconductor. The main idea is that the physical descriptions of the
boundary operator dual to a bulk scalar field can be implemented. Note that we omit the backreaction terms by fixing the metric. 
It is worth noting that in the decoupling limit the backreaction of
the matter fields on the metric is negligible and we set the $U(2)$ charge $q=1$. Here we write the gauge field $A_{\mu}$ in terms of the 
$U(2)$ generators as $A_{\mu}=A^a_{\mu}T_a$ where $T_{0}=\mathbb{I}/2,\,T_{i}=\sigma_{i}/2$ with $\sigma_{i}$ being the Pauli matrices. The field strength tensor, $F_{\mu\nu}=\partial_{\mu}A_{\nu}-\partial_{\nu}A_{\mu}$, is in a common use either in coordinate basis or in the Pauli's frame. The 
 normalized scalar doublet  $\Psi $ can be represented as
\begin{eqnarray}
\Psi =\sqrt{2}\left[
\begin{array}{c}
\lambda(r) \\
\psi (r)
\end{array}%
\right],
\end{eqnarray}
where we have set $\lambda(r)=0$ without loss of
generality \cite{Amado:2013lia}. Maintaining the symmetry of the metric, we suppose that all functions depend on the radial coordinate  $r$. Here we work in the adapted units where $2\kappa ^{2}=1$ and consider the following consistent ansatz for the fields in our setup \cite{Amado:2013lia}. 
\begin{eqnarray}
A^{(0)}_{0} = \Phi(r),\,\,\,A^{(3)}_{0} = \Theta(r),\,\,\,A^{(1)}_{1} =\omega(r).
\end{eqnarray}
Here all functions are real-value and all other fields in Eq.(\ref{action}) are set to zero. The set of the equations of motion (EoM) for scalar and gauge fields is the 
 following \cite{Amado:2013lia}
\begin{eqnarray}
&&\psi ^{\prime \prime }+\Big(\frac{f^{\prime
}}{f}+\frac{2}{r}\Big)\psi
^{\prime }+\Big(\frac{(\Phi -\Theta )^{2}}{4f^{2}}-\frac{m^{2}}{f}-\frac{%
	\omega^{2}}{4r^{2}f}\Big)\psi =0 \,, \label{eq1} \\&&
\Phi ^{\prime \prime }+\frac{2\Phi ^{\prime }}{r}-\frac{\psi
	^{2}}{f}(\Phi
-\Theta ) =0 \,, \label{eq2} \\&&
\Theta ^{\prime \prime }+\frac{2}{r}\Theta ^{\prime }+\frac{\psi ^{2}}{f}%
(\Phi -\Theta )-\frac{\omega^{2}}{r^{2}f}\Theta = 0 \,, \label{eq3} \\&&
\omega^{\prime \prime }+\frac{f^{\prime }}{f}\omega^{\prime }+\Big(\frac{\Theta ^{2}}{f^{2}}%
-\frac{\psi ^{2}}{f}\Big)\omega =0.  \label{eq4}
\end{eqnarray}%
Note that only non vanishing  component of the doublet scalar condensate field, i.e. $\psi(r)$ appeared in the EoMs and here the 
prime means the derivative with respect to the $r$. Phase transitions in the contexts of AdS/CFT was studied numerically in Ref.\cite{Amado:2013lia} and analytically in Ref.\cite{Momeni:2013bca} (see also many aspects of this mixed model in \cite{Mukhopadhyay:2018oju}-\cite{Nie:2014qma}).

The normal phase of the mixing system occurs when the temperature 
 $T>T_{c}$, and can be described by the exact solutions obtained in Eqs.(\ref{eq1}-\ref{eq4}) with the following prescribed set of functions:
\begin{eqnarray}
f&=&r^{2}-\frac{r_{+}^{3}}{{r}},\,\,\psi =\omega=0,\\\Theta &=&\mu_3\,\Big(1-\frac{1}{r}%
\Big),\,\,\Phi =\mu \Big(1-\frac{1}{r}\Big).
\end{eqnarray}
The balanced holographic superconductor can be achieved by setting 
$\mu_3 = 0$. If we keep $\mu_3\neq 0$, we achieve the unbalanced regime. The asymptotic forms of the solutions near the AdS boundary $r\to\infty$ are 
\begin{eqnarray}
\psi &\cong& \frac{<\mathcal{O}_{+}>}{r^{\Delta _{+}}}+\frac{<\mathcal{O}_{-}>%
}{r^{\Delta _{-}}}\,,\\
\omega&\cong &\frac{<J_{x}^{1}>}{r}+<J_{x}^{0}>\,,\\
\Phi &=&\mu -\frac{\rho }{r}\,,\\
\Theta &=&\mu_3-\frac{\rho _3}{r}\,,
\end{eqnarray}%
where, in the terminology of AdS/CFT,  $\Delta _{\pm }\geq 1$ is the conformal dimension and the  scalar doublet mass $m^{2}=-2> 
m_{BF}^{2}=-\frac{9}{2}$ \cite{BF1,BF2}. Note that on the dual side, $\mu$ is the chemical potential corresponding to the overall
$U(1)\subset U(2)$ generated by $T_{0}$, whereas $\mu_{3}$ is the chemical potential corresponding to the $U(1)\subset SU(2)$ generated by $T_{3}$. Furthermore the expectation value of the dual boundary operator $\mathcal{O}_{\pm }$ plays the role of
the scalar order parameter. Here we have also a current operator
$J_{x}$ as the source term of the AdS asymptotic solution of $w$.

Making the radial coordinate as a compact parameter in a bounded interval, we set the horizon radius $r_{+}=1$ and change to the new coordinate $z=\frac{1}{r}$. As a result, the EoMs (\ref{eq1}-\ref{eq4}) take the following
forms:
\begin{eqnarray}
&&\psi ^{\prime \prime }-\frac{(2+z^{3})}{z(1-z^{3})}\psi ^{\prime }+\Big[%
\frac{(\Phi -\Theta )^{2}}{4(1-z^{3})^{2}}+\frac{2}{z^{2}(1-z^{3})}-\frac{%
	w^{2}}{4(1-z^{3})^{2}}\Big]\psi =0\,,  \label{eq11} \\
&&\Phi ^{\prime \prime }-\frac{\psi ^{2}}{z^{2}(1-z^{3})}(\Phi -\Theta
) =0 \,,
\label{eq22} \\
&&\Theta ^{\prime \prime }+\frac{\psi ^{2}(\Phi -\Theta )}{z^{2}(1-z^{3})}-%
\frac{w^{2}\Theta }{1-z^{3}} =0 \,, \label{eq33} \\
&&\omega^{\prime \prime }-\frac{3z^{2}}{1-z^{3}}\omega^{\prime }+\Big[\frac{\Theta ^{2}}{%
	(1-z^{3})^{2}}-\frac{\psi ^{2}}{z^{2}(1-z^{3})}\Big]\omega =0.
\label{eq44}
\end{eqnarray}%
The aim of the next section is to show that a condensate field $\psi$ in the phase portrait undergoes chaotic paths and furthermore study the three-dimensional trajectories of $\{z,\psi,\psi'\}$. We will develop a numerical code for both a  balanced case as well as an unbalanced case. Furthermore we will investigate how the chaotic phase in the potential field $\Phi$ depends on $z$ and will quantify a possible criticality in the gauge field value $\Theta$ in the vicinity of the horizon of the AdS black holes. Note that the horizon in the new coordinate $z$ is mapped to the $z=1$.

\section{Chaos in balanced  holographic superconductors }
\label{3}
In  this section, we show that the chaotic behavior can be generated because of chaotic structure of the condensate field $\psi$ given in Eqs.(\ref{eq11}-\ref{eq44}) for both balanced and unbalanced cases. The strategy is to write and develop a higher-order algorithm using \textsc{Matlab \& simulink} to solve the system of the EoMs given in Eqs.(\ref{eq11}-\ref{eq44}). In the first attempt, we solve the equations in the functional form $f=f(z)$ by selecting a value of chemical potential $\mu$ in the balanced case such that $\mu_3=0$. Note that the numerical solutions are convergent only for two values given by $\frac{1}{\mu}=10,50$. We need to specify boundary conditions (BCs) of the set of functions in Eqs.(\ref{eq11}-\ref{eq44}). The BCs are given as field values of the AdS boundary when $z\to0$. In this case we observe that the scalar field $\psi$ vanishes at the boundary point $\psi(z\to 0)=0$. It is remarkable to emphasis the role of the AdS boundary of the information which we can get from the boundary values of the distinct fields. At AdS boundary, we utilize 
\begin{eqnarray}
\psi(z\to 0) &\cong& \langle\hat{\mathcal{O}}_{+}\rangle z^{\Delta _{+}},\\
\omega(z\to 0)&\cong &\langle\hat{J}_{x}^{0}\rangle,\\
\Phi(z\to 0) &=&\mu \,,\\
\Theta(z\to 0) &=&0\,,
\end{eqnarray}%
where we have supposed that in the balanced system $\mu_3=0$. A damping form for scalar field $\psi$ can be detected when we make numerical integration. Note that the values of $\langle O_+\rangle$ and $\langle O_-\rangle$ contain key information in AdS/CFT, and differ for different orbits of $\Psi$ but will have a universal form in terms of temperature. In our numerical investigation the initial values are given by the values of $\langle O_+\rangle,\,\langle \hat{J}_{x}^{0}\rangle$ and $\mu$. The sensitivity of the chaotic phase portrait for the scalar condensate field can be demonstrated using a measure functional, in analogue to the Lyapononv measure. By performing a numerical integration, we can demonstrate that $\Psi$ and $\Psi'$ near the boundary always tend to zero and we can take this asymptotic value as a fixed point. Although we observe that the above “fixed point” $\Psi(r)$ always goes to zero at AdS boundary because of the power law dependence of $z^{\Delta _{+}}$ of the field and hence it doesn't mean that  $\langle\hat{\mathcal{O}}_{+}\rangle$ must vanish. Basically as we know from the numerical computations of the  Eqs.(\ref{eq11}-\ref{eq44}) in Ref.\cite{Amado:2013lia}, $\langle\hat{\mathcal{O}}_{+}\rangle$ is independent on the spatial coordinate $z$ and is given as $\langle\hat{\mathcal{O}}_{+}\rangle\sim \sqrt{1-(\frac{T_c}{T})^\delta}$ where $\delta$ is the critical exponent of the condensation phase. In the single $s$ or $p$ phases, we have $\delta=2$; while in the mixing phase $1<\delta<2$.

\begin{figure}[H]
	\centering
	\includegraphics[width=0.44\textwidth]{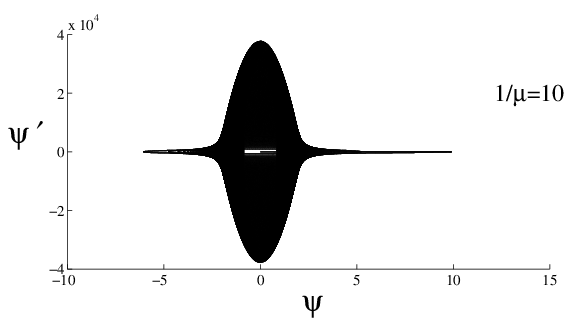}
     \includegraphics[width=0.55\textwidth]{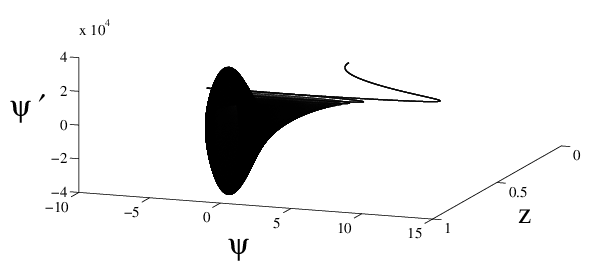}
	\caption{Plots show the chaotic phase portrait $\Psi-\Psi'$ in balanced phase for $1/\mu=10$ (left panel) and their corresponding three-dimensional trajectory $(z,\psi,\psi')$ (right panel). }\label{fig:stream}
\end{figure}

The numerical solutions for the condensate are presented in the Fig.(\ref{fig:stream}). The lower panels show the three-dimensional solutions in an extended phase space spanned by $\{z,\psi,\psi'\}$. We observe that the system initialized from a given initial phase space point $(\psi(z_0),\psi'(z_0))$ is close to the fixed points which undergoes a very chaotic and sensitive pattern. As long as we get towards the black hole horizon $z\sim 1$, the condensate field grows fast and meanwhile the gradient of the condensate field  $\psi'$ also increases  monotonically but non periodic. It is remarkable to mention here that near the horizon the pattern of the phase portrait as illustrated on top panels is completely chaotic. The chaotic phase is defined as regions of the phase space at which the Poincare section is covered mostly by ergodic chaos patterns. A completely phase portrait shows that our dynamical system generically is is covered mostly by ergodic chaos patterns. Condensation field is a chaotic field and it shows that in the balanced case the system dictates chaos. Furthermore, using bifurcation theory, we claim that the condensation field as an order parameter plays the role of the bifurcation parameter. The dynamics of the whole system starting from a single phase is drastically bifurcated near the horizon when the condensate field grows up limitlessly. Although the bifurcation model is handled numerically, we have a type of one-to-many saddle point bifurcation phenomena. More concretely, we run the code for two values of the chemical potential. Note that in both cases, the chaotic pattern remains the same. This is a scale invariant chaotic pattern where the scaling behaviors do not affect the pattern of the dynamics. 
\begin{figure}[H]
	\centering
     \includegraphics[width=0.44\textwidth]{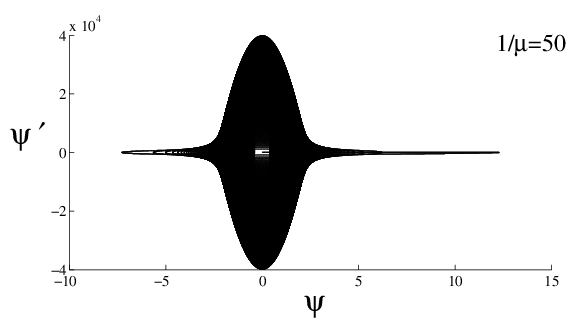}
     \includegraphics[width=0.55\textwidth]{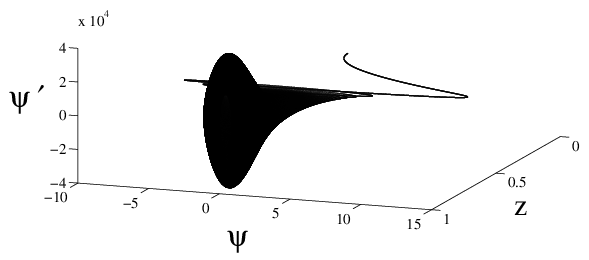}
	\caption{Plots show the chaotic phase portrait $\Psi-\Psi'$ in balanced phase for $1/\mu=50$ (left panel) and their corresponding three-dimensional trajectory $(z,\psi,\psi')$ (right panel). }\label{fig:stream0}
\end{figure}
\begin{figure}[H]
	\centering
    \includegraphics[width=0.5\textwidth]{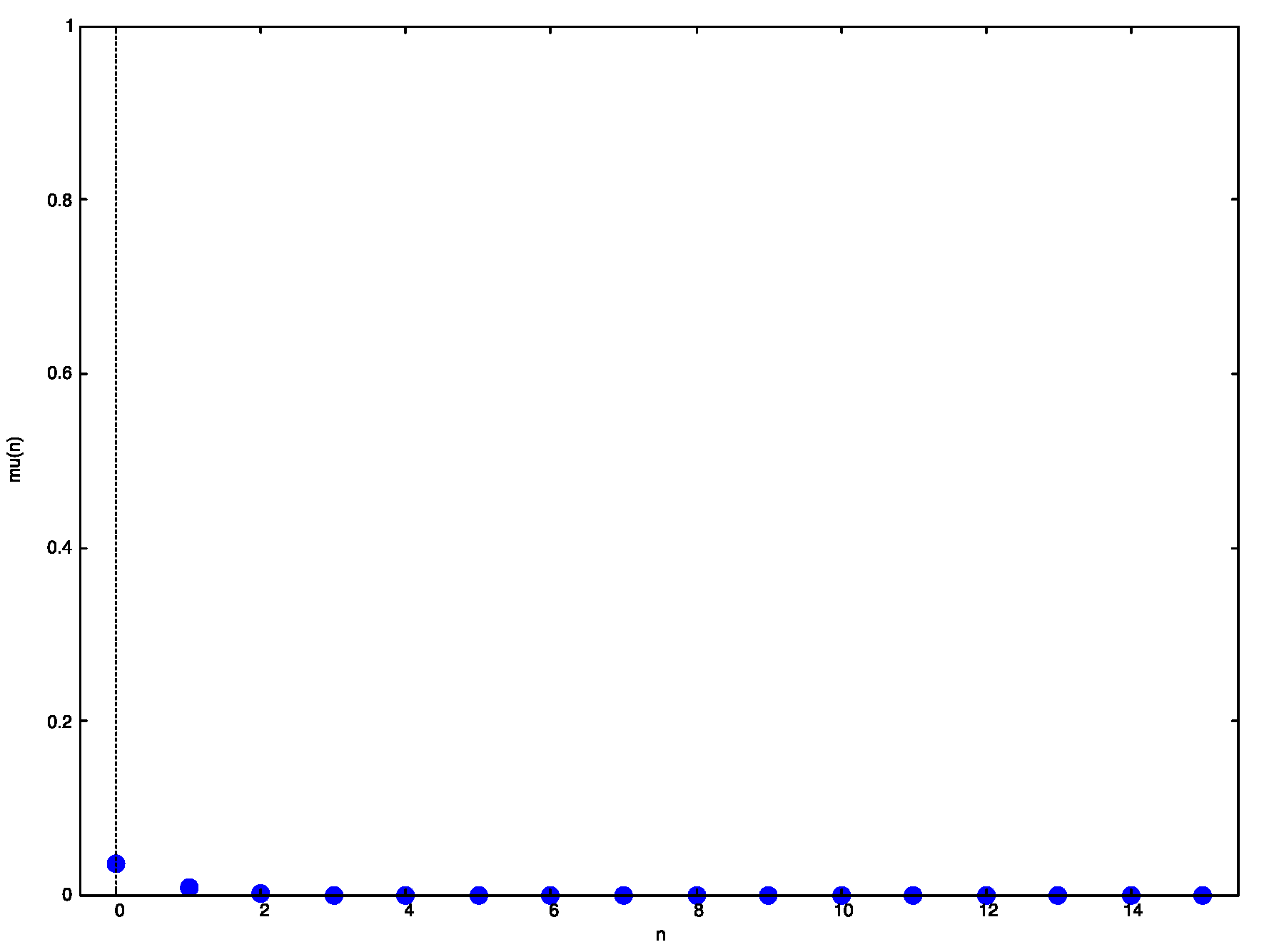}
	\caption{We plot a measure for chaos, i.e. Eq.(\ref{mu}) using a \textsc{Monte Carlo Method} for simulation performed over a range of $n$. We observe that for $n\leq 2$ the measure $\tilde{\mu}(\psi,\psi'|n)>0$. It provides a very powerful evidence to prove the existence of chaotic dynamics in the system of $s+p$ when we consider the order parameter $\psi$ as chaotic field. }\label{measure-balanced}
\end{figure}
\par
Although we have a chaotic phase space for the condensate field in the balanced case, it is reasonable to study more chaos situation in this model. The tool which will be useful is to define a well-defined and enough smooth measure to quantify and prove the existence of chaos in our complicated, non linear dynamical system. There are different ways to detect chaos in nonlinear dynamical systems mainly in the models with a gauge/gravity dual, e.g. the Sachdev-Ye-Kitaev model. Very recently it has been shown that this model results in a quantum chaos behavior near the criticality \cite{Garcia-Garcia:2017bkg} using out-of-time-order four-point correlation function, as a measure for chaotic behavior. This function  controls small quantum  corrections, and it has been demonstrated using this measure that black holes are quantum chaotic systems out of equilibrium \cite{sekino}. In connection to the AdS/CFT proposal, in the bulk portrait, when the criticality appears in the AdS blackhole sector, in Ref.\cite{Chabab:2018lzf} chaotic behavior in the phase space plane can be demonstrated using Poincare-Melnikov theory. The technique here is very similar to Lyapunov function where we can investigate the stability of a given fixed point.  We will search for poles in complex plane of  the relevant Melnikov function. The next step toward discovering chaos is to identify the  type of chaos either temporal or spatial. A simple strategy is to find the homoclinic orbits. In our system of the equations describing superfluity, we need another equivalent strategy by working through the EoMs given in Eqs.(\ref{eq11}-\ref{eq44}). However, there, it is not possible to define Melnikov function. In the context of AdS/CFT, a bound on the Lyapunov exponent constrains the space of putative holographic CFTs with a well defined AdS gravity dual \cite{Perlmutter:2016pkf}. Furthermore it has been shown that the chaotic phase  appears only at an intermediate energy density scale \cite{Hashimoto:2016wme}. The measure of the chaos is provided by the Lyapunov exponent  which is defined as the distance between the two $z$-evolution orbits of solutions. Along the Lyapunov exponent, we define 
\begin{equation}
  \tilde{\mu}(n)=\lim_{\substack{z \to 0 \\ d_0\to 0}} \frac{1}{z}\ln\frac{d(z)}{d_0}
\end{equation}
where $d(z)$ is the distance between the two $z$-evolution orbits of the condensate field $\psi(z)$. The difficulty is how to define a "well-posed" $d(z)$. Any functional form $d(z)=d(\psi,\psi'|n)$ satisfying a set of mathematical requirements can use as the Lyapunov exponent. Let us try to explore a good measure function. It is required that $\tilde{\mu}(n)>0$ and becomes small at $z\to 0$ if the system is chaotic. 

We use a novel definition of the measure, recently used to quantify chaos in  dynamical systems as a generalization of the standard measure defined in \cite{li}. For our dynamical system (randomly distributed over a specifically finite interval $I=(0,1)$), a measure of chaos is defined as follows:
\begin{eqnarray}
\tilde{\mu}(\psi,\psi'|n)=\sup _{z_{<},z_{>}}\int_{0}^{1}\Big[F^{*}_{z_{<},z_{>}}(y,\psi,\psi'|n))-F_{z_{<},z_{>}}(y,\psi,\psi'|n))
\Big]dy, \{z_{<},z_{>}\}=^{\sup}_{\inf}(y\in I)\label{mu},
\end{eqnarray}
with $0\leq F_{z_{<},z_{>}}(y,\psi,\psi'|n))\leq F^{*}_{z_{<},z_{>}}(y,\psi,\psi'|n))\leq 1$. Note that the measure $\tilde{\mu}(\psi,\psi'|n)$ should be monotonically nonnegative. A sufficient condition to have chaotic behavior in the system is that $\tilde{\mu}(\psi,\psi'|n)>0$. In Fig.(\ref{measure-balanced}), we plot the measure for the balanced case using $n$ as chaotic control variable in the measure function. We perform \textsc{Monte Carlo Method} code to estimate a measure function, for a very wide range of the condensate field values $\psi$ as chaotic field. 
We observe that the measure $\tilde{\mu}(\psi,\psi'|n)$ is non negative showing the existence of the chaotic dynamics in the condensation dynamics. 

\section{Chaos in unbalanced holographic superconductors}
\label{4}
\begin{figure}[H]
	\centering
	\includegraphics[width=0.35\textwidth]{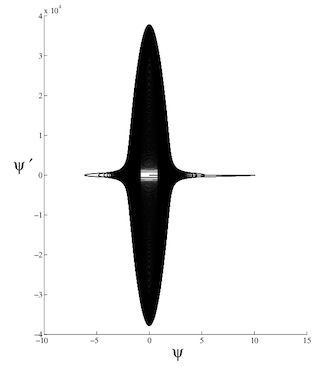}
    \includegraphics[width=0.5\textwidth]{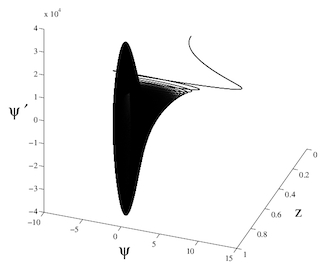}
	\caption{Plots show the chaotic phase portrait $\Psi-\Psi'$ in unbalanced phase for $\frac{1}{\mu}=10,\,-\frac{\mu_3}{\mu}=-0.1$ (left panel) and their corresponding three-dimensional trajectory $(z,\psi,\psi')$ (right panel).}
	\label{fig:stream11}
\end{figure}

\begin{figure}[H]
	\centering
    	\includegraphics[width=0.35\textwidth]{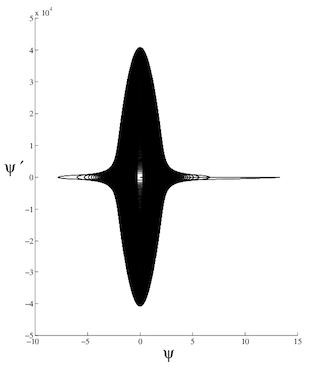}
        \includegraphics[width=0.5\textwidth]{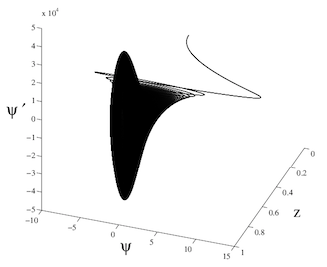}
	\caption{Plots show the chaotic phase portrait $\Psi-\Psi'$ in unbalanced phase for $\frac{1}{\mu}=10,\,-\frac{\mu_3}{\mu}=-0.9$ (left panel) and their corresponding three-dimensional trajectory $(z,\psi,\psi')$ (right panel).}
	\label{fig:stream12}
\end{figure}

\begin{figure}[H]
	\centering
        \includegraphics[width=0.35\textwidth]{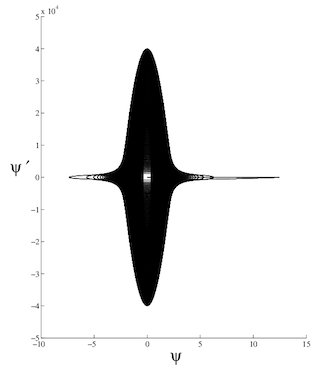}
        \includegraphics[width=0.5\textwidth]{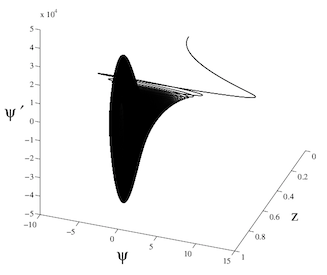}
	\caption{Plots show the chaotic phase portrait $\Psi-\Psi'$ in unbalanced phase for $\frac{1}{\mu}=50,-\frac{\mu_3}{\mu}=-0.1$ (left panel) and their corresponding three-dimensional trajectory $(z,\psi,\psi')$ (right panel).}
	\label{fig:stream13}
\end{figure}

\begin{figure}[H]
	\centering
          \includegraphics[width=0.35\textwidth]{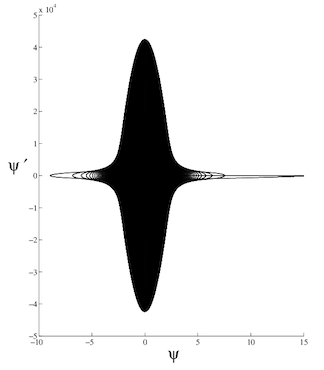}
         \includegraphics[width=0.5\textwidth]{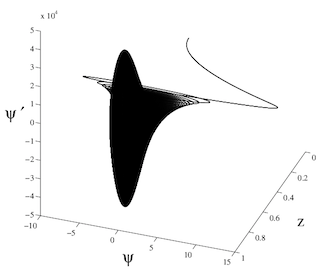}
	\caption{Plots show the chaotic phase portrait $\Psi-\Psi'$ in unbalanced phase for $\frac{1}{\mu}=50,-\frac{\mu_3}{\mu}=-0.9$ (left panel) and their corresponding three-dimensional trajectory $(z,\psi,\psi')$ (right panel).}
	\label{fig:stream14}
\end{figure}

\begin{figure}[H]
	\centering
	\includegraphics[width=.45\textwidth]{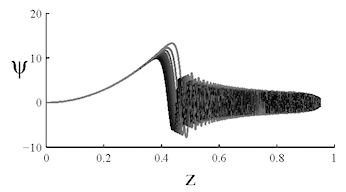}
    \includegraphics[width=.45\textwidth]{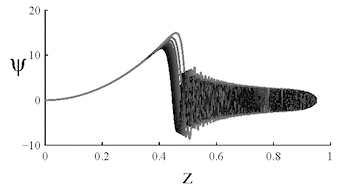}
     \includegraphics[width=.45\textwidth]{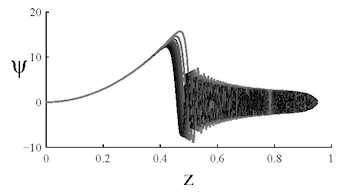}
	\caption{Plots show numerical solution for  $\psi(z)$ with $\frac{1}{\mu}=10$ (upper left), $\frac{1}{\mu}=30$ (upper right) and $\frac{1}{\mu}=50$ (lower) in unbalanced phases. }
	\label{fig:stream3}
\end{figure}

\begin{figure}[H]
	\centering
	\includegraphics[width=.45\textwidth]{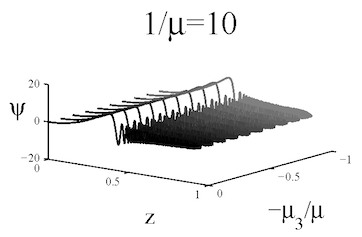}
    \includegraphics[width=.45\textwidth]{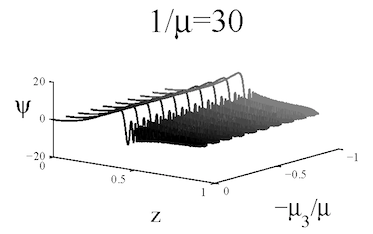}
     \includegraphics[width=.45\textwidth]{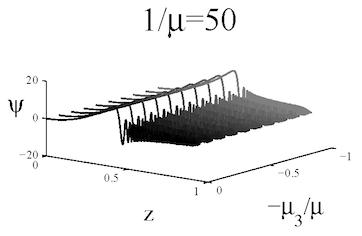}
	\caption{Plots show numerical solution for three-dimensional trajectory $(-\frac{\mu_3}{\mu},z,\psi)$ with $\frac{1}{\mu}=10$ (upper left), $\frac{1}{\mu}=30$ (upper right) and $\frac{1}{\mu}=50$ (lower) in unbalanced phases. }
	\label{fig:stream4}
\end{figure}

\begin{figure}[H]
	\centering
	\includegraphics[width=.45\textwidth]{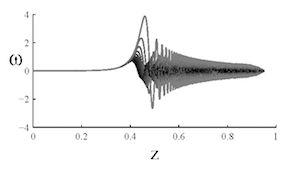}
    \includegraphics[width=.45\textwidth]{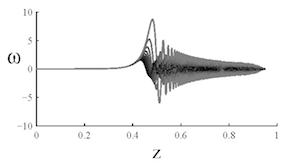}
    \includegraphics[width=.45\textwidth]{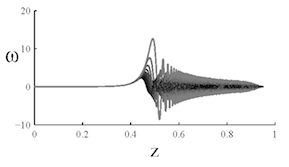}
	\caption{Plots show numerical solutions for  $\omega(z)$ with $\frac{1}{\mu}=10$ (upper left), $\frac{1}{\mu}=30$ (upper right) and $\frac{1}{\mu}=50$ (lower) in unbalanced phases.}
	\label{fig:stream5}
\end{figure}

\begin{figure}[H]
	\centering
	\includegraphics[width=.45\textwidth]{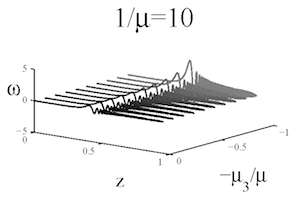}
    \includegraphics[width=.45\textwidth]{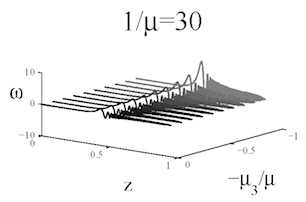}
    \includegraphics[width=.45\textwidth]{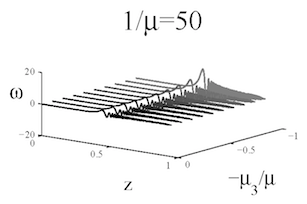}
	\caption{Plots show numerical solutions for three-dimensional trajectory $(-\frac{\mu_3}{\mu},z,\omega)$ with $\frac{1}{\mu}=10$ (upper left), $\frac{1}{\mu}=30$ (upper right) and $\frac{1}{\mu}=50$ (lower) in unbalanced phases.}
	\label{fig:stream6}
\end{figure}

\begin{figure}[H]
	\centering
\includegraphics[width=0.7\textwidth]{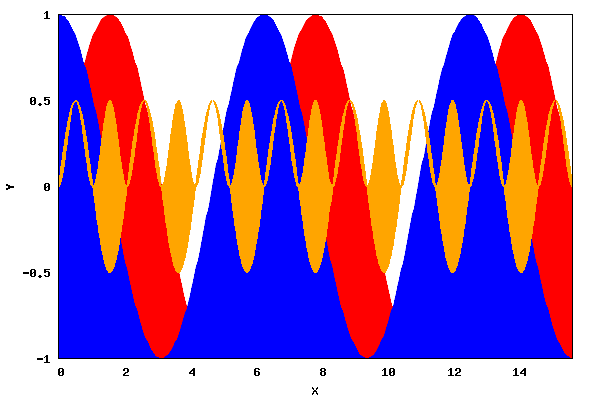}
	\caption{The measures as given in Eqs.(\ref{mu2},\ref{mu3}) v.s. $x=\{n,m\}$ as chaotic control variables. The blue region corresponds to the (\ref{mu2}) and red is for (\ref{mu3}). Note that there is a common range for both $x=\{n,m\}$, shown with orange where for the system, both measures  $Y=\{\tilde{\mu}_1(\psi,\psi'|n),\tilde{\mu}_2(\omega,\omega'|m))\}>1$, showing a chaotic dynamics in a system.}
	\label{mu1mu2}
\end{figure}

In the previous section, we demonstrate that chaos is probably the dominant dynamical behavior of the system. However, a type of the saddle point is still hard to quantify. Furthermore we introduce a very useful measure for chaos introduced in EQ.(\ref{mu}). Using the positivity of this measure, we have a trustable evidence for chaos. In the previous section, we worked only on the balanced superconductors where the dual chemical potential is restricted to just one variable $\mu$. The $s+p$ superconductor beyond the balanced regime is also very interesting to investigate and in this section we investigate such systems. Note that the phase portrait and measure function provide a viable strategy to examine chaotic dynamics in the system. We develop a \textsc{Matlab CODE} to solve the system of the EoMs given in Eqs.(\ref{eq11}-\ref{eq44}) by imposing a suitable set of the boundary conditions on the fields. Our numerical results are presented in Figs.(\ref{fig:stream11}-\ref{mu1mu2}). For each graph, we plot the three-dimensional trajectory in an extended phase portrait for $\{z,f(z),f'(z)\}$, where $f(z)$ stands for the condensate field $\psi(z)$. In Figs.(\ref{fig:stream11}-\ref{fig:stream6}), we  plotted   $(z,\psi(z),\psi'(z))$, for potential values ranged from $\mu^{-1}=10,\,30,\,50$ and $\frac{\mu_3}{\mu}=0.1-0.9$. In all cases, the system starts from an initial point at very tiny $z$, corresponding to a point near the AdS boundary. Notice that the field vanishes near the boundary point. This behavior results from the inverse power-law decay of the field from the field theory point of view. The condensate field $\psi$ and its gradient $\psi'(z)$ are slowly growing up with a smooth slope till the vicinity of the black hole horizon, and exactly very close to the horizon of the black hole. The amplitude of the condensate field  decreases when the symmetry breaking phenomena occurs, but the gradient $\psi'(z)$ eventually undergoes a resonant-like phenomena. At the vicinity of the horizon, very close to $z\sim 1$, there is a singular point, $\lim (\psi,\psi')|_{z\to 1^{+}}\to 0$. The limit circle around the saddle point $z=1$, black hole horizon, is an evidence for chaotic behavior. Furthermore, the system bifurcates from one saddle point at $z\to 0$ to a instable saddle point. Consequently we can state that the system will start from a large distance half-unstable saddle point  located at the AdS boundary  $z\to 0_{-}$(there is no more physical space far from the AdS boundary point $z\to 0^{+}$), and finally ends to half unstable saddle point near the horizon of the black hole (here there is no region of space with $z\to 1^{-}$). 

It is remarkable to discuss the variation of the condensation field $\psi(z)$ also as a single valued function. The essence and origin of the saddle points singularities and their roles in the critical chaotic deterministic dynamics of the system are clarified and displayed in Fig.(\ref{fig:stream4}) and (\ref{fig:stream6}), where we plotted for $(-\frac{\mu_3}{\mu},z,\psi)$. Note that when the ratio of chemical potentials increases at a fixed value of $-\frac{\mu_3}{\mu}$, the space coordinate variation of $\psi(z)$ behaves as a damping of a non linear oscillator, i.e., the condensate field $\psi$ for a whole region from AdS boundary point $z\sim 0^{-}$ to a region close to the black hole horizon at the peak near $z=0.4$. An exponential decay of the field continues to the black hole horizon. It shows that the decay is exponential and the saddle point $z=1^{-}$ is a half stable point. A remarkable note is that the system is still unstable near horizon because of the large momentum transfer (large $\psi'(z)\gg 1$ ) to the system. 
 
Furthermore clearly a similar chaotic-like behavior appears in the Fig.(\ref{fig:stream5}) for the field $\omega(z)$ in the doublet field $\Psi$. We detect  the phase space at which the Poincare section is covered mostly by many ergodic chaos patterns in Fig.(\ref{fig:stream11}-\ref{fig:stream14}). For the field, $\omega$, we made plots  using $(-\frac{\mu_3}{\mu},z,\psi)$. When the ratio of the chemical potentials increases at a fixed value of the $-\frac{\mu_3}{\mu}$, the space coordinate variation of $\omega(z)$ behaves like a super-damping non linear oscillator. It oscillates even more than the condensate field $\psi$ for a domain in the region from AdS boundary point $z\sim 0_{-}$ to a region close to the black hole horizon at the peak near $z=0.4$. A speedily exponential decay of the adjoint field $\omega(z)$ continues to the black hole horizon. It shows that the decay is exponential and the saddle point $z=1^{-}$ is also a half stable point.
It is reasonable to estimate the measure of chaos $\mu$ given in Eq.(\ref{mu}). Note that the measure can be estimated in a similar code as we have used in the balanced system. However, the measure function can be different when considering an unbalanced case since we have two dual potential variables $\mu, \mu_3$. Furthermore, the chaos can also show up in $\omega$ as well as $\psi$. For our new dynamical system (again randomly distributed over a specifically finite interval $I=(0,1)$), a measure of chaos is defined for $\psi$ as follows:
\begin{eqnarray}\label{mu2}
\tilde{\mu}_1(\psi,\psi'|n)=\sup _{z_{<},z_{>}}\int_{0}^{1}\Big[F^{*}_{z_{<},z_{>}}(y,\psi,\psi'|n))-F_{z_{<},z_{>}}(y,\psi,\psi'|n))
\Big]dy\,,
\end{eqnarray}
with $\{z_{<},z_{>}\}=^{\sup}_{\inf}(y\in I)$ and $0\leq F_{z_{<},z_{>}}(y,\psi,\psi'|n))\leq F^{*}_{z_{<},z_{>}}(y,\psi,\psi'|n))\leq 1$, and for $\omega$
\begin{eqnarray}\label{mu3}
\tilde{\mu}_2(\omega,\omega'|m)=\sup _{z_{<},z_{>}}\int_{0}^{1}\Big[F^{*}_{z_{<},z_{>}}(y,\omega,\omega'|m))-F_{z_{<},z_{>}}(y,\omega,\omega'|m))
\Big]dy\,,
\end{eqnarray}
with $\{z_{<},z_{>}\}=^{\sup}_{\inf}(y\in I)$ and $0\leq F_{z_{<},z_{>}}(y,\omega,\omega'|n))\leq F^{*}_{z_{<},z_{>}}(y,\omega,\omega'|m))\leq 1$. The measure functions $\tilde{\mu}_1(\psi,\psi'|n),\tilde{\mu}_2(\omega,\omega'|m)$  should be  monotonically  nonnegative. The enough and sufficient condition to have chaotic behavior in the system is that $\tilde{\mu}_1(\psi,\psi'|n)>1~\&~\tilde{\mu}_2(\omega,\omega'|m)>1$ at least for an interval $(n,m)$. Furthermore  it is required to have  $\tilde{\mu}_1(\psi,\psi'|n)\,\&\,\tilde{\mu}_2(\psi,\psi'|m)>0$ becomes small at $z\to 0$ if the system is chaotic.
In Fig.(\ref{mu1mu2}), we plot the measure for the balanced case using $x=\{n,m\}$ as chaotic control variables in the measure function. We perform \textsc{Monte Carlo Method} code to estimate measure functions, for a very wide range of the condensate field values $\psi,\omega$ as chaotic fields. 
As we observed, there are common domains for $x=\{n,m\}$, where the measured pair functions $Y=\{\tilde{\mu}_1(\psi,\psi'|n),\tilde{\mu}_2(\omega,\omega'|m))\}$ are non negative. Therefore it shows the existence of the chaotic dynamics in the condensation dynamics. 



\section{Conclusion}
\label{con}
Chaos is defined as the stochastic behavior occurring in a deterministic dynamical  system \cite{def}. When the dynamical systems have many degrees of freedom as well as highly nonlinearity, it is possible to have significantly drastic and globally  sensitive dynamical evolutions for a tiny change in the initial conditions. Remarkably when the system of our study contains phase transitions from normal phase to the super fluidity in type II superconductors, the probability of chaotic evolution highly increases. Using gauge/gravity duality (or AdS/CFT), a toy model for mixture of superconductors with both competitive $s,p$ modes can be implemented. It was shown that the model undergoes phase transitions in a wide range of the phases as well as different values for chemical potentials. In this work, we examined the $s+p$ chaotic point of view where we looked for some evidences to demonstrate the existence of chaotic dynamics explicitly. As an attempt to quantify the chaos, we investigated the phase portrait for the condensate field $\psi$ for different values of the chemical potential. A numerical integration of the set for the coupled non linear differential equations showed that the phase portrait starts at the AdS boundary, ends to a chaotic behavior and resonances as a damping mode near the black hole horizon. Because the horizon is the location for the criticality in the system, we demonstrated that the criticality happens not only at the certain temperature range and in a specific critical temperature depending on the parameters of the model but also when the system have chaotic behavior. 

Chaos here is a universal behavior because of its independence on the values of the dual chemical potential $\mu$ as well as other parameters. Remarkably, it is depended on the initial values of the field at the AdS boundary, very far from the horizon. In order to ensure the chaotic form of the condensate, we also focused on the measure function for chaos: a one parameter (in the balanced case) function of the condensate field and its gradient and by performing MC simulation. We discovered that the measure functional was a positive definite for a range of parameters and a sample of random field variables. Such analysis was repeated for the unbalanced case, where both of the scalar doublet condensate fields show a chaotic behavior not only in the phase space but also using the two-parameter functions of the chaos. Furthermore we argued that this super fluidity can be described in language of chaos and fractal geometry. The fractal geometry emerged from this mixed system will provide a better description for the chaotic behavior in the system. In the context of AdS/CFT,  a bound on the Lyapunov exponent constrains the space of putative holographic CFTs with a well defined  AdS gravity dual \cite{Perlmutter:2016pkf}. Furthermore it has been shown that the chaotic phase  appears only at an intermediate energy density scale \cite{Hashimoto:2016wme}. The key to understand the physics of chaos in holographic superconductors is to compute  the spectral functions of the dual holographic  superconductor in normal states and compare them with the boundary theory, mainly as the characteristic of the marginal Fermi liquid. Very recently a report showed that the fermion condensate undergoes a chaotic behavior and it demonstrated that  the admixture subsystem is responsible for such a chaotic behavior \cite{zverof}. Using AdS/CFT we can estimate  the spectral functions. We argue here that the energy spectrum at ground state close to the $T = 0$  is nondispersive. The reason of chaos in the bulk theory is the existence of an  admixture subsystem in bulk, which is here a component of the $SU(2)$ gauge field. Further studies need to be done to compute spectral functions for mixed system and study their zero temperature limit.
\section*{Acknowledgements}
D. Momeni \& M. Al Ajmi would like to acknowledge the support of Sultan Qaboos University under the Internal Grant (IG/SCI/PHYS/19/02).
PC is financially supported by the Institute for the Promotion of Teaching Science and Technology (IPST) under the project of the \lq\lq Research Fund for DPST Graduate with First Placement\rq\rq\,, under Grant No.\,033/2557.


\end{document}